\documentclass[aps,prl,twocolumn,amsmat,amssymb,amsfonts,superscriptaddress]{revtex4-1}
\usepackage{amsmath}
\usepackage{times}
\usepackage[varg]{txfonts}
\usepackage{hyperref}
\usepackage{pdfpages}

\usepackage{color}
\usepackage{graphics}
\usepackage{graphicx}
\usepackage[percent]{overpic}
\usepackage{wasysym}
\usepackage{caption}
\usepackage{subcaption}

\newcommand{\avg}[1]{\langle #1 \rangle}

\newcommand{\h}[1]{{#1}^{\dagger}} 

\newcommand{\cb}[1]{\bar{#1}}

\newcommand{\up}{\uparrow}
\newcommand{\down}{\downarrow}

\newcommand{\del} {\partial}

\newcommand{\nairo}{Na${}_2$IrO${}_3$ }
\newcommand{\liiro}{Li${}_2$IrO${}_3$}
\newcommand{\airo}{A${}_2$IrO${}_3$ }

\definecolor{acolor1}{RGB}{68,34,221}
\definecolor{acolor2}{RGB}{28,181,0}
\definecolor{acolor3}{RGB}{238,34,34}

\hypersetup{colorlinks=true,linkcolor=acolor1,citecolor=acolor1,urlcolor=acolor1} 
\begin{document}

% Title of paper
\title{Generic Spin Model for the Honeycomb Iridates beyond the Kitaev Limit}
%Authors and affiliation
\author{Jeffrey G. Rau}
\author{Eric Kin-Ho Lee}
\affiliation{Department of Physics, University of Toronto, Toronto, Ontario M5S 1A7, Canada}
\author{Hae-Young Kee}
\email[Electronic Address: ]{hykee@physics.utoronto.ca}
\affiliation{Department of Physics, University of Toronto, Toronto, Ontario M5S 1A7, Canada}
\affiliation{Canadian Institute for Advanced Research/Quantum Materials Program, Toronto, Ontario MSG 1Z8, Canada}

\date{\today}
 \begin{abstract}
Recently, realizations of Kitaev physics have been sought in the \airo family of honeycomb
iridates, originating from oxygen-mediated exchange through edge-shared octahedra. However, for the $j_{\rm eff} = 1/2$ Mott insulator in these materials exchange from direct $d$-orbital overlap
is relevant, and it was proposed that a Heisenberg term should be added to the Kitaev model. 
Here we provide the generic nearest-neighbour spin Hamiltonian when both
oxygen-mediated and direct overlap are present, containing a bond-dependent off-diagonal exchange in addition to Heisenberg and Kitaev terms. We analyze this complete model using a combination of
classical techniques and exact diagonalization. Near the Kitaev limit, we find new magnetic phases, 120${}^\circ$ and incommensurate spiral order, as well as extended regions of zigzag and stripy order. Possible applications to \nairo and \liiro\ are discussed.
\end{abstract}
\maketitle

The honeycomb family of iridium oxides\cite{singh2010antiferromagnetic,liu2011long,singh2012relevance,ye2012direct,lovesey2012magnetic,choi2012spin,comin2012na_,clancy2012spin,gretarsson2013magnetic,gretarsson2013crystal,cao2013evolution} has attracted a considerable
amount of attention
\cite{shitade2009quantum,chaloupka2010kitaev,kimchi2011kitaev,bhattacharjee2012spin,mazin2012na_,kim2012topological,kim2012topological,chaloupka2013zigzag,okamoto2013global,foyevtsova2013ab} 
due to the possibility they lie
near a realization of Kitaev's exactly solvable spin-1/2
honeycomb model\cite{kitaev2006anyons}. This
model hosts a number of remarkable features: a
$Z_2$ spin liquid with gapless Majorana fermions and (non-Abelian) anyonic excitations under an applied
magnetic field. No symmetry principle excludes
terms besides the Kitaev, so additional
interactions are generically expected.
 From microscopic calculations of exchange
mediated through the edge-shared oxygen octahedra,
it has been proposed that a pure Kitaev model of $j_{\rm eff}=1/2$ spins was the appropriate description\cite{jackeli2009mott}.
It was further suggested that direct overlap of the $d$-orbitals generalizes this to a Heisenberg-Kitaev (HK) model\cite{chaloupka2010kitaev},
linearly interpolating between an isotropic
Heisenberg model and Kitaev's bond-dependent exchange Hamiltonian. 
Extensive study of the HK model\cite{jiang2011possible,reuther2011finite,trousselet2011effects,schaffer2012quantum,price2012critical,price2013finite} %,kimchi2013kitaev}
has shown a variety of fascinating
phenomena, including an extended spin liquid phase and quantum phase transitions
into several well-understood magnetic ground states. 
While present, the zigzag phase seen in \nairo\cite{liu2011long,choi2012spin,ye2012direct} is difficult to stabilize within
the HK model; one must resort to additional $t_{2g}$-$e_g$ exchange paths\cite{chaloupka2013zigzag}
or further neighbour hoppings\cite{kimchi2011kitaev}. 
In light of this puzzle one may question whether the HK model provides an adequate
description of the honeycomb iridates even at the nearest neighbour level.

In this Letter, we show that when applied
to the honeycomb iridates the HK
model is incomplete, explicitly deriving the $j_{\rm eff}=1/2$ spin model from a multiorbital $t_{2g}$
Hubbard-Kanamori Hamiltonian.
Considering the most idealized
crystal structure, an additional
spin-spin interaction beyond the HK model must be included: bond-dependent
symmetric off-diagonal exchange. 
The complete spin Hamiltonian has the form
\begin{equation}
  \label{hkg-model}
  H = \sum_{\avg{ij} \in \alpha\beta(\gamma)} \left[J \vec{S}_i\cdot \vec{S}_j + K S^\gamma_i S^\gamma_j +
    \Gamma \left(S^\alpha_i S^\beta_j + S^\beta_i S^\alpha_j\right)\right],
\end{equation}
where $J$ is Heisenberg exchange, $K$ is the Kitaev exchange,
and $\Gamma$ denotes the symmetric off-diagonal exchange.
On each bond we distinguish one spin direction $\gamma$, labeling the bond
$\alpha\beta(\gamma)$ where $\alpha$ and $\beta$ are the two remaining directions.
Examining the phase diagram using a combination of classical arguments and exact diagonalization,
we find that with the inclusion of $\Gamma$ new magnetic phases are stabilized near the Kitaev limits: an incommensurate spiral (IS) and 120${}^\circ$ order,
in addition to extended regions of zigzag and stripy order.

\begin{figure}[tp]
  \includegraphics[width=0.85\columnwidth]{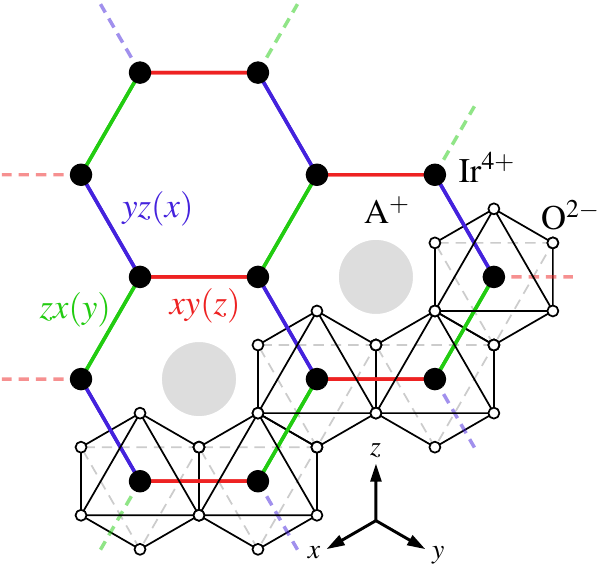}
\caption{
\label{fig:exchanges}Crystal structure of the honeycomb iridates ${\rm A}_2{\rm Ir}{\rm O}_3$
with ${\rm Ir}^{4+}$ in black, ${\rm O}^{2-}$ in white, and ${\rm A} = {\rm Na}^{+},{\rm Li}^{+}$ in gray.
For the Kitaev and bond-dependent exchanges
we have denoted the $yz(x)$ bonds blue,
the $zx(y)$ bonds green and the $xy(z)$ bonds red.}
\end{figure}

\emph{Microscopics}.-- 
We first construct a minimal model of
a honeycomb lattice of Ir$^{4+}$ ions surrounded by a network of
edge-sharing oxygen octahedra.
The ${\rm Ir}^{4+}$ $5d$ levels are split into an $e_g$ doublet
and $t_{2g}$ triplet by large crystal field effects, leaving a single hole in the $t_{2g}$ states. 
Within the $t_{2g}$ manifold, the orbital angular momentum behaves as an $l_{\rm eff} = 1$ triplet, with large
spin-orbit coupling splitting this into an active $j_{\rm eff}=1/2$ doublet and filled $j_{\rm eff}=3/2$ states. Because of
significant on-site interactions, localized
$j_{\rm eff}=1/2$ spins provide an effective model for the low-energy physics.
To perform the strong coupling expansion, we consider an atomic
Hamiltonian of Kanamori form\cite{sugano1970multiplets}:
\begin{equation}
  \label{eq:atomic}
  H_0 = \sum_i\left[ \frac{U-3 J_H}{2} (N_i-5)^2 - 2J_H S_i^2 -\frac{J_H}{2} L_i^2\right],
\end{equation}
where $N_i$, $S_i$, and $L_i$ are the total number, spin, and (effective) orbital
angular momentum operators at site $i$, $U$ is the Coulomb interaction, and $J_H$ is Hund's coupling.
The expansion is carried out
in the limit $U,J_H \gg \lambda \gg t$, first taking $U$ and $J_H$ to be large. Since
the spin-orbit coupling then dominates the kinetic terms, the resulting spin-orbital
model can be projected into the $j_{\rm eff}=1/2$ subspace.

\begin{figure}[tp]
  \begin{subfigure}[b]{0.95\columnwidth} 
    \begin{overpic}[width=\textwidth]%,grid,tics=10]        
      {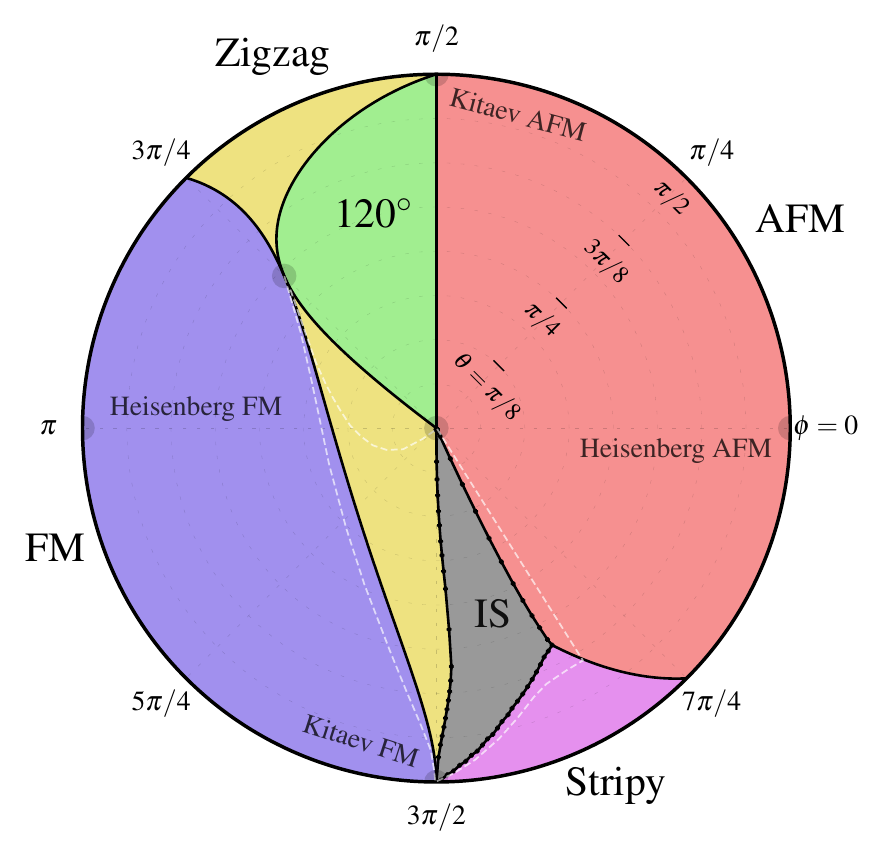}
    \end{overpic}
    \caption{\label{fig:cl-pd} Classical phase diagram with $\Gamma>0$}
  \end{subfigure}
  \begin{subfigure}[b]{0.75\columnwidth} 
    \begin{subfigure}[t]{0.32\columnwidth} 
      \includegraphics[width=\textwidth]{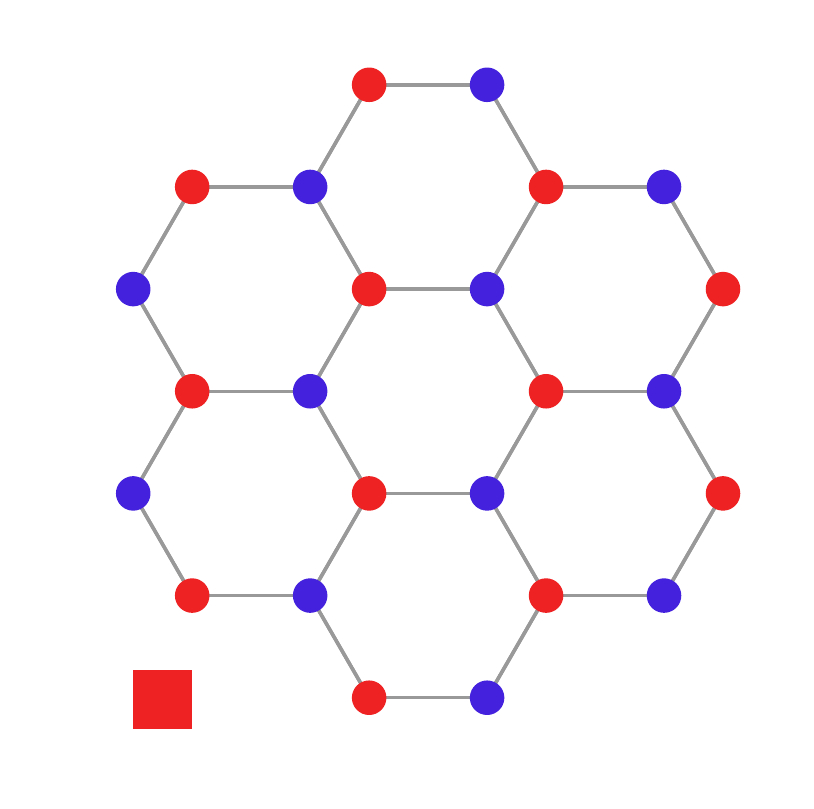}
      \caption{\label{fig:cl-afm}AFM }
    \end{subfigure}
    \begin{subfigure}[t]{0.32\columnwidth} 
      \includegraphics[width=\textwidth]{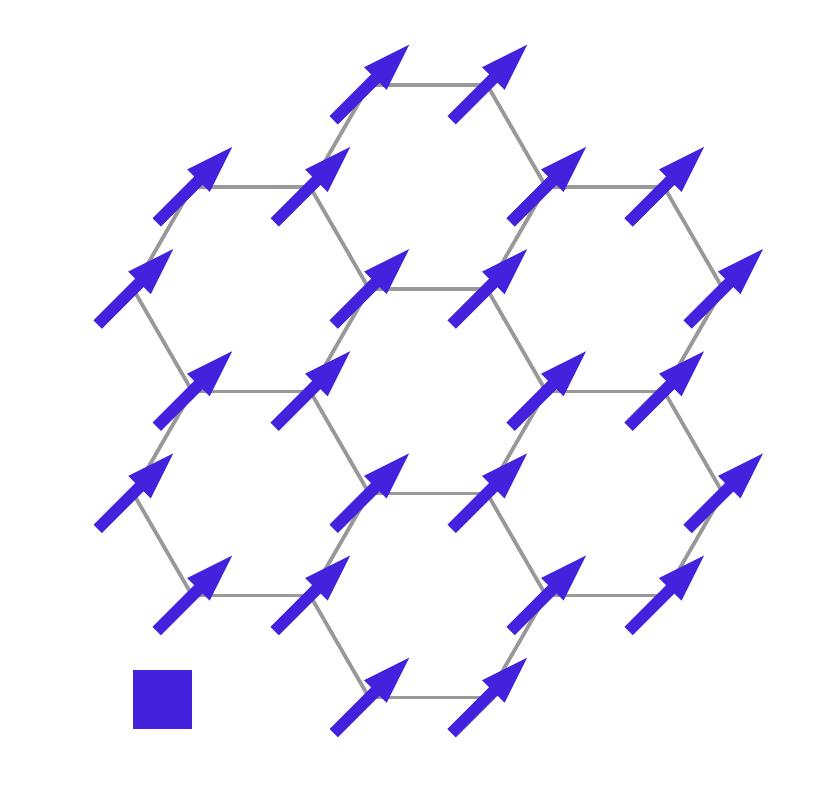}
      \caption{\label{fig:cl-fm}FM }
    \end{subfigure}
    \begin{subfigure}[t]{0.32\columnwidth} 
      \includegraphics[width=\textwidth]{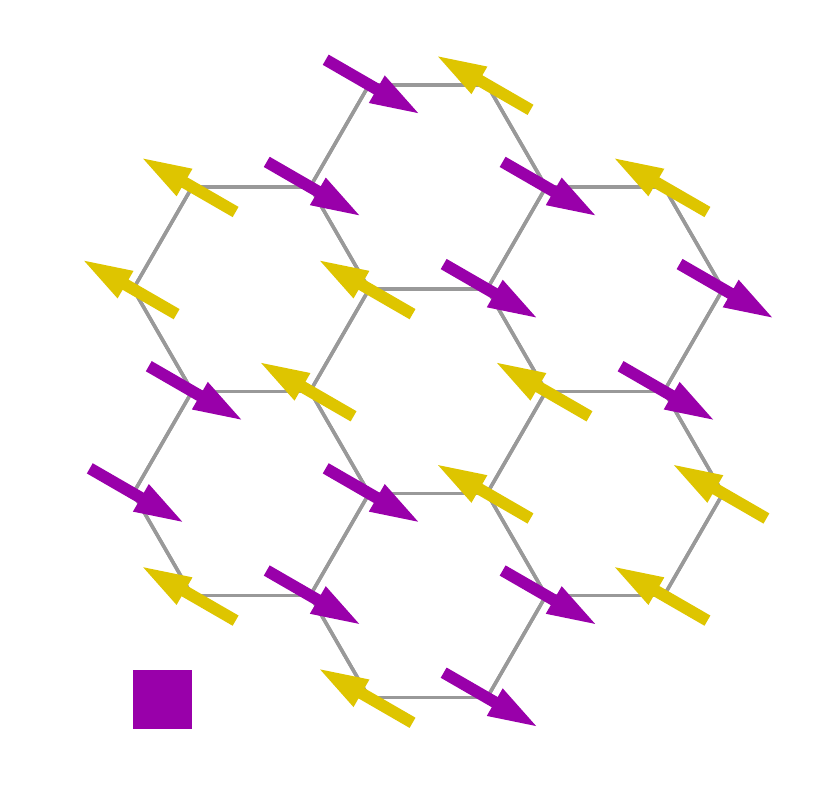}
      \caption{\label{fig:cl-stripy}Stripy }
    \end{subfigure}
    \begin{subfigure}[t]{0.32\columnwidth} 
      \includegraphics[width=\textwidth]{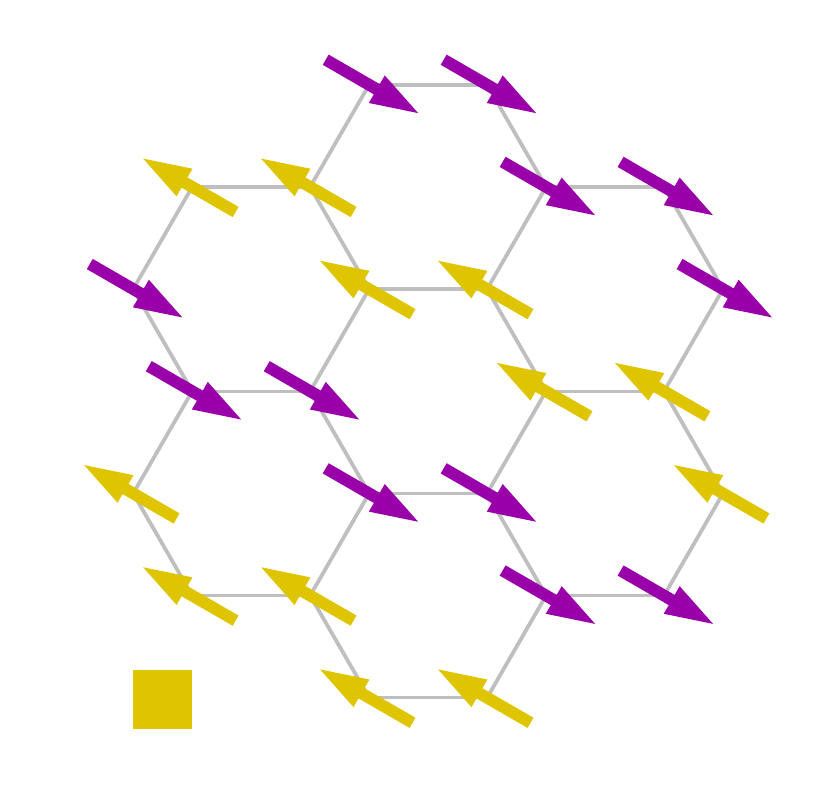}
      \caption{\label{fig:cl-zigzag}Zigzag }
    \end{subfigure}
    \begin{subfigure}[t]{0.32\columnwidth} 
      \includegraphics[width=\textwidth]{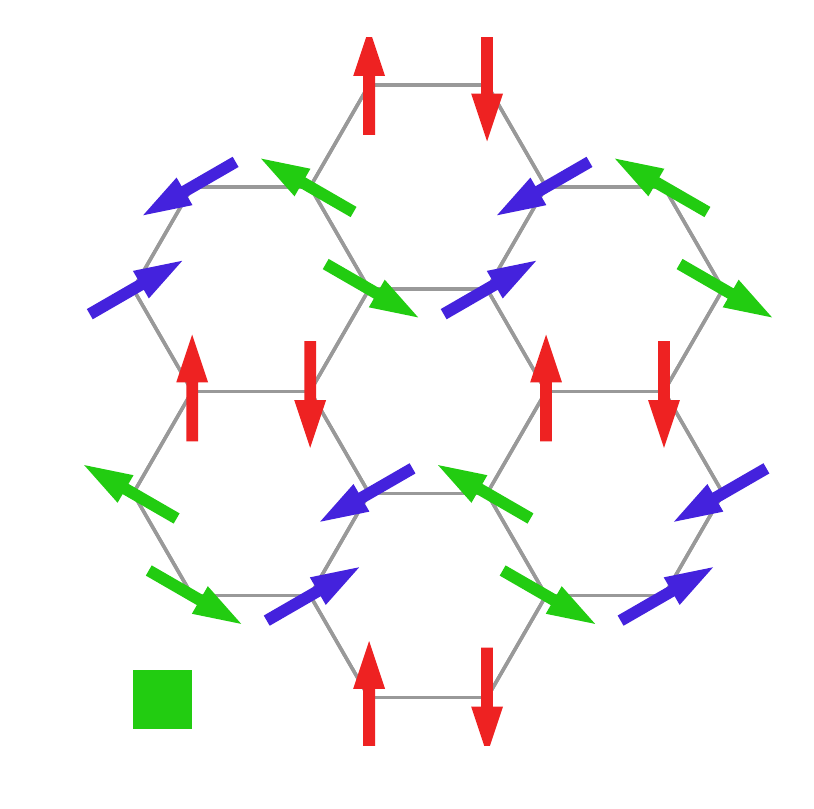}
      \caption{\label{fig:cl-120}120$^\circ$ }
    \end{subfigure}
  \end{subfigure}  
  \begin{subfigure}[t]{0.24\columnwidth} 
    \includegraphics[width=\textwidth]{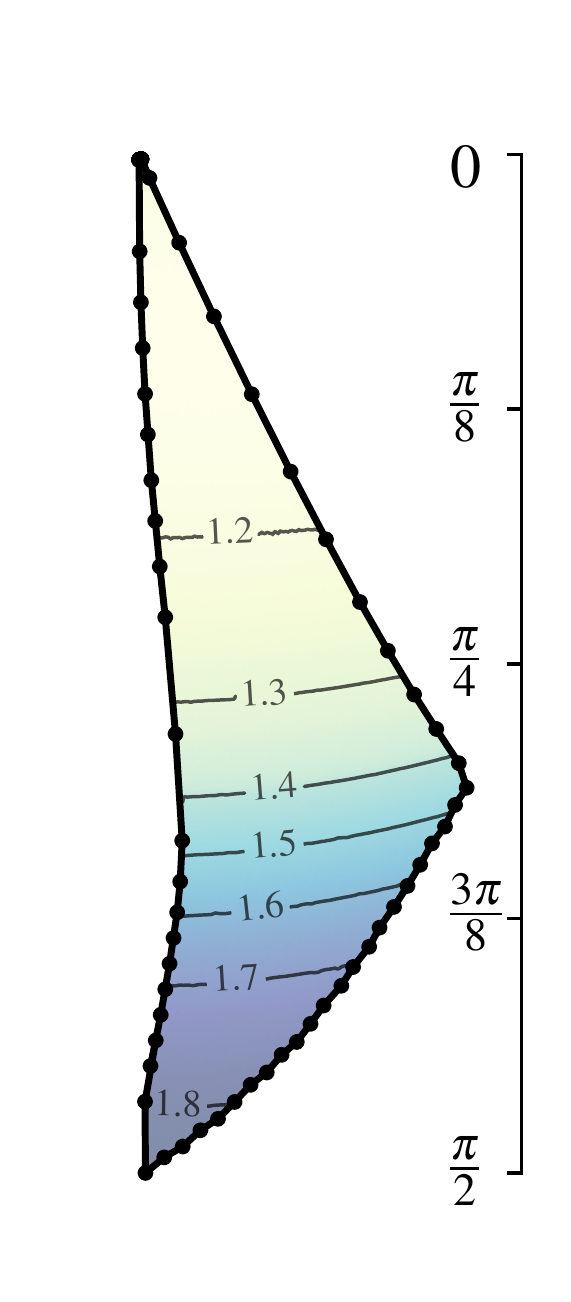}
    \caption{\label{fig:cl-is}$|\vec{Q}|$ in the IS}
  \end{subfigure}
\caption{
\label{fig:classical-phase-diagram}
(a) Combined Luttinger-Tisza and single-$Q$ analysis. Solid colours correspond
to exact classical ground states from Luttinger-Tisza while the region indicated by
the white dashed line are the single-$Q$ results. (b-f) Ground state spin configurations in each phase. (g) Magnitude of
the ordering wave-vector $\vec{Q}$ in the IS phase.
}
\end{figure}

The kinetic terms are encapsulated through a tight-binding model for the $\rm Ir$ $t_{2g}$ orbitals,
including both direct overlap of $d$-orbitals and hopping mediated through the oxygen atoms. 
For our purposes, we focus on nearest-neighbour bonds where we then have
\begin{equation*}
  \sum_{\avg{ij}\in \alpha\beta(\gamma)} \left[
    t_1\left(\h{d}_{i\alpha} d_{j\alpha}+\h{d}_{i\beta} d_{j\beta}\right)+
    t_2\left(\h{d}_{i\alpha} d_{j\beta}+\h{d}_{i\beta} d_{j\alpha}\right)+
    t_3 \h{d}_{i\gamma} d_{i\gamma}
    \right],
\end{equation*}
where $\h{d}_{i\alpha} = (\h{d}_{i\alpha \up}\ \h{d}_{i\alpha \down})$ and $d_{i\alpha}$ are the creation and annihilation operators
for the $t_{2g}$ state $\alpha$ at site $i$. Here we sum over the $yz(x)$, $zx(y)$ and $xy(z)$
links as indicated in Fig. \ref{fig:exchanges}, but mapping the directions
to orbitals as $x \rightarrow yz$, $y \rightarrow zx$ and $z \rightarrow xy$.
The parameters $t_1$, $t_2$, and $t_3$ are given by
\begin{eqnarray*}
  t_1 = \frac{t_{dd\pi}+t_{dd\delta}}{2},  \ \
  t_2 = \frac{t_{pd\pi}^2}{\Delta_{pd}} +\frac{t_{dd\pi}-t_{dd\delta}}{2},  \ \
  t_3 = \frac{3 t_{dd\sigma} +t_{dd\delta}}{4},
\end{eqnarray*}
where $t_{dd\sigma}$, $t_{dd\pi}$, $t_{dd\delta}$ and $t_{pd\pi}$ are Slater-Koster\cite{slater1954simplified}
parameters for the direct ${\rm Ir}$-${\rm Ir}$ overlap and ${\rm Ir}$-${\rm O}$
overlap while $\Delta_{pd}$ is the ${\rm Ir}$-${\rm O}$ gap\footnote{
An additional $xy-xz$ and $xy-yz$ hopping is allowed
by the full crystal symmetry, but is not present when we consider low-order processes involving only
two neighbouring Ir atoms with ideal oxygen octahedra.
}. Treating the kinetic terms as a perturbation yields the Hamiltonian
in Eq. \ref{hkg-model} with
\begin{eqnarray}
  \label{eq:exchanges}
  J &=& \frac{4}{27}\left[
    \frac{
      6t_1(t_1+2t_3) 
    }{U-3J_H}
    +
    \frac{2( t_1-t_3)^2}{U-J_H}
    +
    \frac{
      (2t_1+t_3)^2
    }{U+2 J_H}
    \right],
    \\
  K &=& \frac{8 J_H}{9}\left[
    \frac{(t_1 - t_3)^2-3t_2^2}{(U-3J_H)(U-J_H)}
    \right],
    \\
  \Gamma &=& \frac{16 J_H}{9} \left[
    \frac{t_2(t_1-t_3)}{(U-3J_H)(U-J_H)}
  \right].
\end{eqnarray}
Exchanges of the same form as the $\Gamma$ term were originally
called symmetric anisotropic exchange\cite{moriya1960anisotropic,dzyaloshinsky1958thermodynamic}
and can be related to the truncated dipolar exchange\cite{gardner2010magnetic,vanrynbach2010orbital} discussed in other contexts through a reparametrization. 
We stress that since this term is allowed by symmetry even in the most idealized cases,
the presence of the $\Gamma$ term is a generic feature of $j_{\rm eff}=1/2$ models
with edge-shared octahedra (see the Supplemental material \cite{supp} for more information).
To confirm this, the strong coupling expansion
was also carried out in the limit where $U,\lambda \gg J_H \gg t$,
with the contributions of $J_H$ included in the excited states perturbatively.
While energies of the virtual states involve $\lambda$
instead of $J_H$, all three terms are generated, with the dependence of $K$ and $\Gamma$ on the hoppings $t_1$, $t_2$, and $t_3$
unchanged (Supplemental Material \cite{supp}).
 Whereas the Kitaev limit can be
naturally accessed when $t_2 \gg t_1,t_3$, leaving this regime introduces both $J$ and $\Gamma$
making it difficult to reach the HK limit\footnote{
Taking $t_2=0$ and $t_1\neq t_3$ gives the HK model, but this corresponds
to the case of corner-shared octahedra, as discussed in \cite{okamoto2013doped}.
Reaching this point in the edge-shared case again requires unrealistic fine-tuning.}.
Fine tuning could in principle render $\Gamma$
small, but the dominant contributions to $t_1 \sim t_{dd\pi}$ and $t_3 \sim t_{dd\sigma}$ are of opposite sign making any such tuning
implausible. Further applications to wider classes of iridium oxides are left for future work.

\emph{Classical phase diagram}.- To understand
the effects of including this bond-dependent $\Gamma$ term, we first
map out the classical magnetic phases.
We parametrize the exchanges using angles $\phi$
and $\theta$ 
\begin{equation}
  J = \sin{\theta} \cos{\phi},\ \ \ \ \
  K = \sin{\theta} \sin{\phi},\ \ \ \ \
  \Gamma = \cos{\theta}, 
\end{equation}
fixing the energy scale so that $\sqrt{J^2+K^2+\Gamma^2} = 1$.
By mapping $\vec{S}_i \rightarrow -\vec{S}_i$ on one sublattice, we send $\phi \rightarrow -\phi$ and $\theta \rightarrow \pi-\theta$,
so we can consider only $\Gamma > 0$.
To obtain the classical phase diagram, the Luttinger-Tisza approximation\cite{luttinger1946theory,litvin1974luttinger} is first used.
In this approximation, the constraint of fixed spin length is released, allowing
for a direct solution of the classical model.
In the regions of the phase diagram where this fails, 
we have further supplemented this with an analysis of a single-$Q$ ansatz.
The combined results are shown in Fig. \ref{fig:cl-pd}
with $0 < \theta \leq \pi/2$ mapped to the radial direction and 
$0 \leq \phi < 2 \pi$ mapped to the angular direction.

When the resulting spin
configuration satisfies the local length constraint,
the Luttinger-Tisza method yields the exact 
classical ground state.  This holds
for most of the phase diagram aside from the region in Fig. \ref{fig:cl-pd}
indicated by dashed white lines.
In this region we consider spin configurations
of the form
\begin{equation}
  \vec{S}_i =  \sin{\eta_i} \left[\hat{e}^x_i \cos{\left(\vec{Q}\cdot\vec{r_i}\right)}+\hat{e}^y_i\sin{\left(\vec{Q}\cdot\vec{r_i}\right)}\right]+\cos{\eta_i} \hat{e}^z_i
\end{equation}
where the canting angles $\eta_i$ and local frames defined by $(\hat{e}^x_i,\hat{e}^y_i,\hat{e}^z_i)$
are independent variational parameters on two sublattices.
The energy of the ansatz is minimized over the variational 
parameters and $\vec{Q}$ for each pair of angles $(\phi,\theta)$.

In the HK
limit [the boundary of the disk in Fig \ref{fig:cl-pd}] there are four classical phases: the ferromagnet (FM),
antiferromagnet (AFM), stripy, and zigzag as
in Figs. \ref{fig:cl-afm}-\ref{fig:cl-zigzag}. These states occupy
large regions of phase space even as $\Gamma$ is introduced,
with the AFM and FM states covering most the phase diagram.
Finite $\Gamma$ breaks the accidental spin rotational symmetry
enjoyed by the FM and AFM states in the (classical) HK\ limit, pinning
the orderings to fixed spatial direction. For $\Gamma>0$, the AFM becomes pinned along
the $[111]$ direction whereas the FM lies in the
plane perpendicular to $[111]$ with all directions degenerate. The stripy and zigzag
phases have the spins in direction $x$, $y$ or $z$ locked to
the orientations of the stripe and zigzag pattern,
tilting slightly away from the stripe and zigzag direction as $\Gamma$ becomes non zero.

The effects of $\Gamma$ are  most evident where a large classical degeneracy is present,
such as near the Kitaev points at $(\phi,\theta)=(\pm\pi/2,\pi/2)$
and near $(0,0)$, where we only have the bond-dependent $\Gamma$ term. Here 
two new states are introduced: 
120$^\circ$ order and an 
incommensurate spiral.
The 120$^\circ$ order with wave vector $\vec{Q} = K$ appears near
the (antiferromagnetic) Kitaev limit at ($\pi/2,\pi/2)$. This is a coplanar spiral,
with the spins lying in the plane perpendicular to $[111]$. The spins are 
at relative angles $0,\pm 2\pi/3$ on the same sublattice (as shown in Fig. \ref{fig:cl-120}), with the relative
angle between sublattices unconstrained.
An additional degenerate point appears at
$(\phi,\theta)=(3\pi/4,\cos^{-1}(\frac{1}{\sqrt{3}}))$ where $J=-K=-\Gamma$,
with the 120$^\circ$, FM, and zigzag phases meeting at a single point\footnote{At this point the spin Hamiltonian takes a truncated dipolar form
$\sim \sum_{\avg{ij}} (\vec{\delta}_{ij}\cdot \vec{S}_i ) (\vec{\delta}_{ij}\cdot \vec{S}_j )$
where $\vec{\delta}_{ij}$ is the bond direction vector.}.
The second large region of
zigzag phase appearing when $\Gamma \gg |J|,|K|$
has the spins predominantly oriented along the $[1\cb{1}\cb{1}]$, $[\cb{1}1\cb{1}]$, and
$[\cb{1}\cb{1}1]$ directions, tilting away slightly as one
explores the phase. The IS phase remains coplanar despite
the $\vec{Q}$ vector varying throughout the phase. The magnitude of the
IS wave vector lies in the range $1.2 < | \vec{Q} | < 1.8$ as
shown in Fig. \ref{fig:cl-is}.

\begin{figure}[!t]
\begin{subfigure}{0.95\columnwidth} 
   \begin{overpic}[width=\textwidth]%,grid,tics=10]  
     {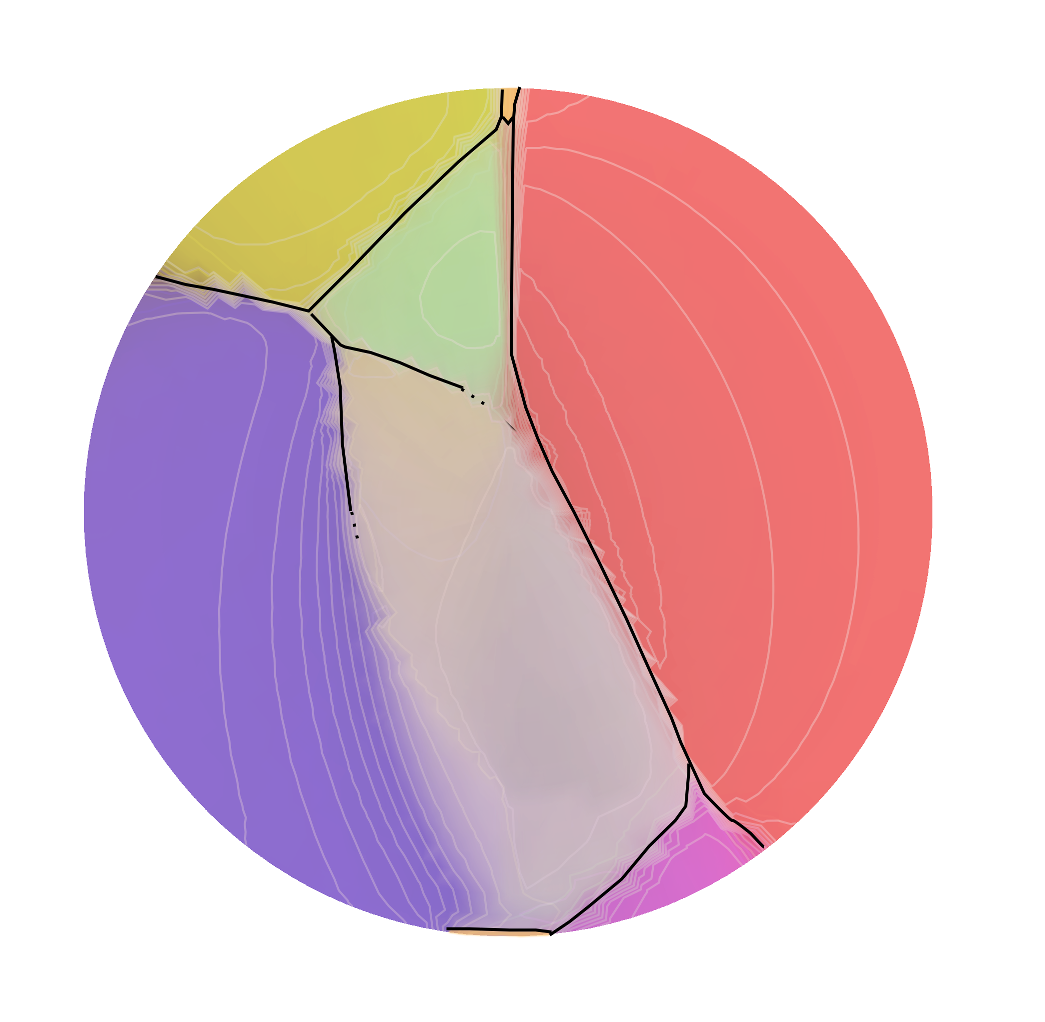}
  \put(-0.85,0){\includegraphics[scale=0.926]{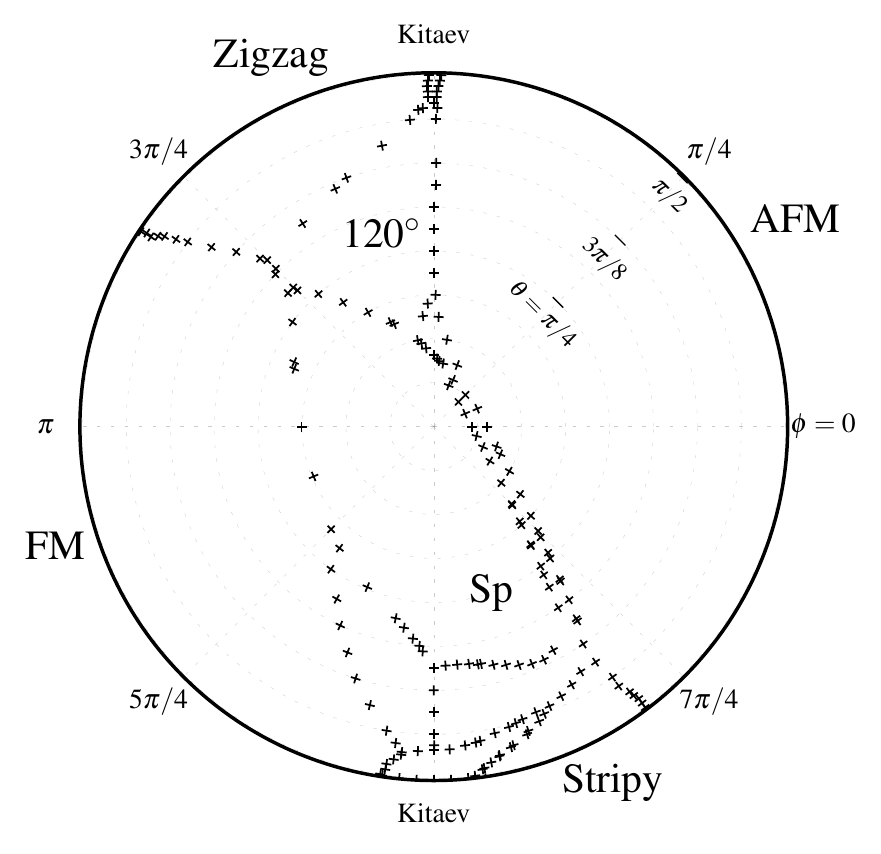}}
  \end{overpic}
    \caption{\label{fig:ed-north}Phase diagram for $\Gamma >0$}
\end{subfigure}
\begin{subfigure}{0.95\columnwidth} 
    \begin{overpic}[width=\textwidth]%,grid,tics=10]        
      {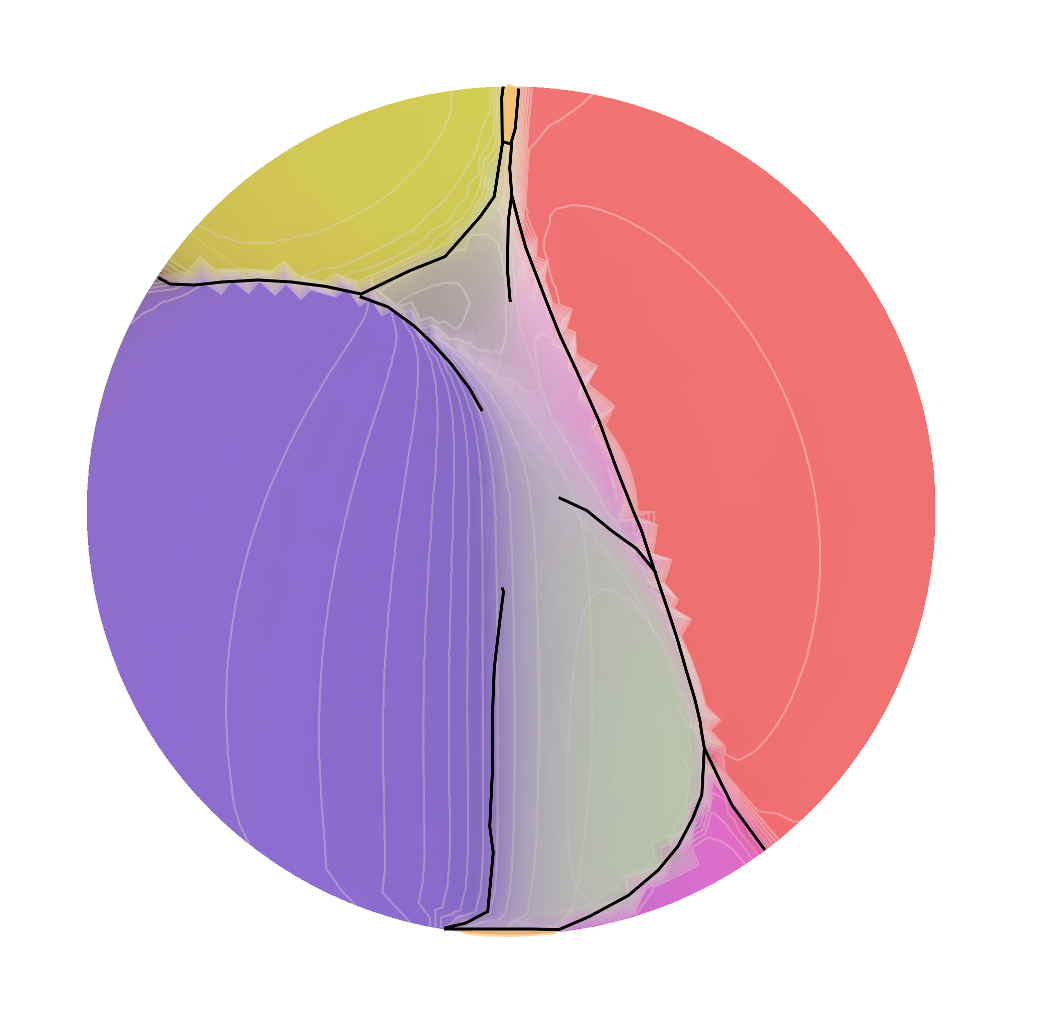}
      \put(-0.85,0){\includegraphics[scale=0.926]{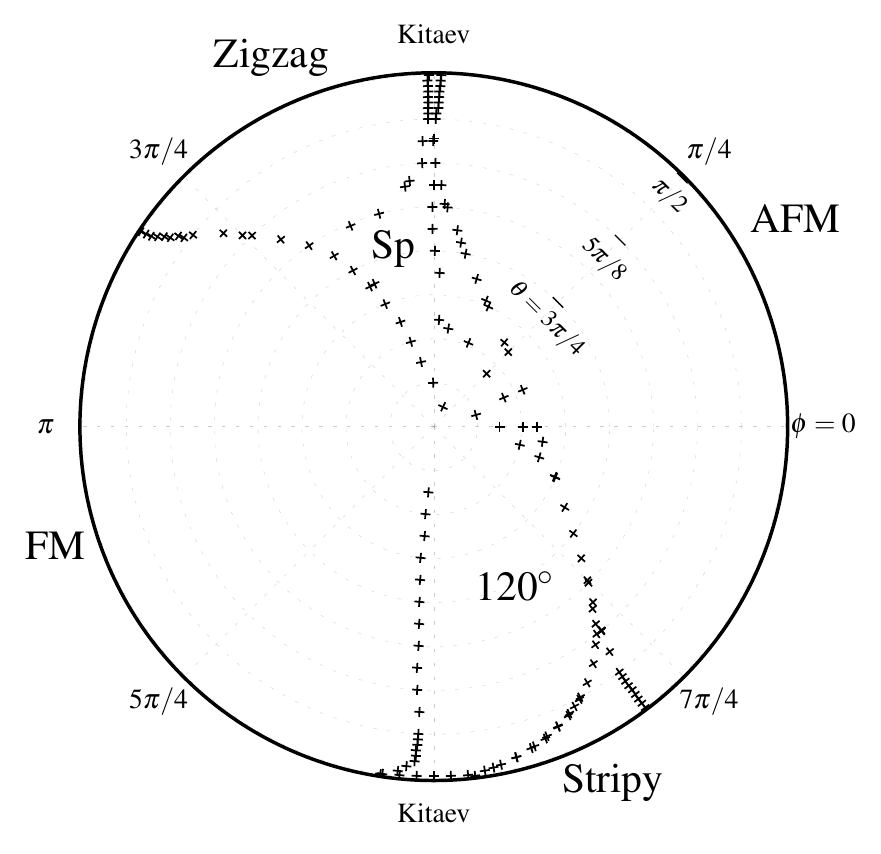}} 
   \end{overpic}
    \caption{\label{fig:ed-south}Phase diagram for $\Gamma <0$}
\end{subfigure}
\caption{\label{fig:ed-phase-diagram}
  [(a),(b)] Phase boundaries from exact diagonalization of a 24-site cluster.
  Markers indicate the location of singular features in $-\del^2 E/\del \phi^2$ or
  $-\del^2 E/\del\theta^2$, with lines to guide
  the eye along the sharp phase boundaries. Colours [as in Fig. \ref{fig:cl-pd}] and contours map magnitudes of the static structure factor.
[as in Eq. \ref{eq:structurefactor}] for each phase. The Kitaev spin
liquid is shown in orange, whereas the spiral phase is shown in dark gray. 
The HK limit lies at the boundary of each disk at $\theta=\pi/2$.
}
\end{figure}

\emph{Exact diagonalization}.- To gain an understanding
of the features of the classical results that carry over to
the full quantum mechanical model we have performed exact diagonalization.
We consider a 24-site cluster that has been used previously to study the HK model\cite{chaloupka2010kitaev,chaloupka2013zigzag,okamoto2013global},
 providing a reasonable description of the phases found at the
classical level as well as the Kitaev spin liquids. 
In the HK limit, the existence of a local spin rotation \cite{khaliullin2005orbital,chaloupka2010kitaev}
that maps $J \rightarrow -J$ and $K \rightarrow K+2J$, gives four well-understood
magnetic limits in addition to the two exactly solvable Kitaev points. These are
the FM, AFM, and their zigzag and stripy images under the mapping.
This transformation is no longer useful as $\Gamma$ is included\footnote{
If one performs the site dependent rotation with the $\Gamma$ term included
the resulting Hamiltonian is not of the same form, with the $\Gamma$ parts taking
on a sign structure with an enlarged unit cell.
}, but the phases surrounding
these points can still be identified with each respective limit.
While the IS phase is unlikely to be well represented 
on such a small cluster, the remaining phases such as the 120$^\circ$ phase
are compatible with the cluster geometry. We note that the transformation
used to relate $\Gamma >0$ to $\Gamma <0$ no longer applies in the
quantum case and so both regions must be analyzed separately.

To identify the phase boundaries, we have computed the second derivatives
of the ground-state energy, $-\del^2 E/\del\phi^2$ and
$-\del^2 E/\del\theta^2$, looking for singular features that indicate changes
in the ground state characteristics. Phases containing
exactly solvable or well-understood points, such as 
the zigzag, stripy, AFM, FM,
and the Kitaev spin liquids can be readily identified. 
The remaining phases were identified by examining the
spin-spin correlation functions $\avg{S^{\alpha}_i S^{\beta}_j}$, primarily through the static structure factor
\begin{equation}
  \label{eq:structurefactor}
S_Q = \frac{1}{N} \sum_{ij} e^{i \vec{Q}\cdot(\vec{r}_i-\vec{r}_j)} \avg{\vec{S}_i\cdot\vec{S}_j}
\end{equation}
in both the original basis and after applying the local spin rotation discussed above\cite{chaloupka2010kitaev}.
The resulting phase diagrams for $\Gamma>0$ and $\Gamma<0$ are presented in Fig. \ref{fig:ed-phase-diagram}, with the
structure factor for each phase plotted using the colours from Fig. \ref{fig:cl-pd} and then overlayed. Contours
indicating lines of constant $S_Q$ in each phase are also shown.
 The phase diagrams bear a remarkable resemblance to the classical results, with the gross
features of the phase diagram preserved for both $\Gamma >0$ and
$\Gamma <0$. 

While the new 120$^\circ$ phase was identified, the results are less
distinctive in the regions where
the Luttinger-Tisza approach failed. Because of the suggestion of 
incommensurate phases from the classical analysis, it is
likely that the small size of cluster used may not properly
capture the behaviour in this region. Nevertheless, 
in the classical IS region exact diagonalization shows a spiral phase (Sp) with
correlations at wave vector $\vec{Q} = K/2$, the
ordering with the longest periodicity allowed by the cluster size. This wave vector
has magnitude $|\vec{Q}| \approx 1.2$, roughly in line with the range prescribed by the
classical calculations. The neighbouring regions are also well defined,
with stripy correlations for $\Gamma<0$ and
zigzag correlations for $\Gamma>0$, as expected from the classical analysis.
While the stripy correlations for $\Gamma<0$ are quite strong, the corresponding
zigzag correlations for $\Gamma>0$ are weak, showing no sharp
transition as one moves into the classical IS region.
At the pure $\Gamma$ limits the correlators become short ranged, with most of the correlators exactly zero. %%
From these results we expect the gross features of the phase diagram to be robust 
to finite size effects except perhaps for the wave vector of the spiral phase.

\emph{Discussion}.- Within the scope of the model presented, the zigzag
phase observed in \nairo\cite{liu2011long,choi2012spin,ye2012direct}
appears only when $J$ is negative. This is plausible: in
Eq. \ref{eq:exchanges} take the Slater-Koster parameters to have the
canonical ratios $t_{dd\sigma} : t_{dd\pi} = 3 : -2$ (assuming that
$t_{dd\delta} \sim 0$) then $2t_1+t_3 \sim 0$ and $t_1(t_1+2t_3)<0$
giving $J<0$ at leading order in $J_H/U$. Additional contributions to
these exchanges, such as on-site oxygen
interactions\cite{chaloupka2010kitaev} and $t_{2g}$-$e_g$
contributions\cite{chaloupka2013zigzag} possibly affect the details.
Further, \emph{ab initio} calculations of the electronic band structure of
\nairo \cite{kim2012topological,mazin2012na_,foyevtsova2013ab} and
\liiro\cite{kim2013strain} suggest that second and third neighbour
hoppings as well as trigonal and other structural distortions may not
be negligible.  Some consequences of further neighbour exchange have
been discussed\cite{kimchi2011kitaev}, but a proper treatment is
missing - one must include the orbital dependence of these hoppings
that results in anisotropic exchanges.  Inclusion of trigonal and
other distortions allows an additional symmetric off-diagonal
exchange\footnote{ The additional symmetric off-diagonal exchange has
  the form $\Gamma' \sum_{\avg{ij} \in \alpha\beta(\gamma)}
  \left(S^\alpha_i S^\gamma_j+S^\gamma_i S^\alpha_j+S^\beta_i
    S^\gamma_j+S^\gamma_i S^\beta_j\right)$ }, but these have been
estimated to be small experimentally\cite{cao2013evolution}. We further
expect that the nearest neighbour model dominates over the longer range 
exchanges, and so including them should not alter the results qualitatively.

We emphasize that understanding the minimal model introduced in this
work is the first step towards a complete picture of the honeycomb
iridates.  Evidence of symmetric off-diagonal exchange can be seen
through anisotropy in the magnetic susceptibility. From a
high-temperature expansion of the model in Eq. \ref{hkg-model}, one
finds $(\Theta_{\perp} - \Theta_{||})/(\Theta_{\perp} + 2\Theta_{||})
=\Gamma / (3J+K)$ independent of $g$-factor anisotropy, where
$\Theta_{||}$ and $\Theta_{\perp}$ are the Curie-Weiss temperatures
for the in- and out-of-plane susceptibilities. Fitting to experimental
data for \nairo\cite{singh2010antiferromagnetic} yields the relation
$\Gamma/(3J+K) \sim -0.3$, showing that if we are near the zigzag
regime where $K \gg |J|$ then there is non-negligible $\Gamma$
exchange.  Given that 120${}^\circ$ and IS order appear in proximity
to the zigzag phase, these could be promising candidates for ordering
in other honeycomb iridates such as \liiro.

\emph{Note added}.- After submission of this work, the existence of the $\Gamma$
term was discussed based on \emph{ab initio} quantum chemistry calculations in Ref. \cite{katukuri2013kitaev}

\emph{Acknowledgements}.-
We thank R. Schaffer, K. Hwang, V. Vijay Shankar and Y. B. Kim for useful discussions.
Computations were performed on the GPC supercomputer at the SciNet HPC Consortium. 
SciNet is funded by: the Canada Foundation for Innovation under the auspices of Compute Canada, 
the Government of Ontario, Ontario Research Fund - Research Excellence; and the University of Toronto.
This work was supported by the NSERC
of Canada, CIFAR, and the Centre for Quantum Materials at the
University of Toronto.

\bibliography{draft}

\newpage



\section*{Supplementary material}

# Supplementary material for "Generic Spin Model for the Honeycomb Iridates beyond the Kitaev Limit"

Jeffrey G. Rau,[1] Eric Kin-Ho Lee,[1] and Hae-Young Kee[1,2,*]

[1]*Department of Physics, University of Toronto, Toronto, Ontario M5S 1A7, Canada*

[2]*Canadian Institute for Advanced Research/Quantum Materials Program, Toronto, Ontario MSG 1Z8, Canada*

## I. TIGHT-BINDING MODEL FOR EDGE-SHARING OCTAHEDRA

Consider a single *xy(z)* type bond, placing the Ir$^{4+}$ ions at the origin and along $\hat{y} - \hat{x}$ as shown in Fig. 1. The remaining bonds can generated using lattice symmetries. We denote the corresponding $t_{2g}$ electron operators as $d_1^\dagger = (d_{1,yz}^\dagger\ d_{1,zx}^\dagger\ d_{1,xy}^\dagger)$ and $d_2^\dagger = (d_{2,yz}^\dagger\ d_{2,zx}^\dagger\ d_{2,xy}^\dagger)$. The bond Hamiltonian can be seen to be

$$T + T^\dagger = d_1^\dagger T_{12} d_2 + d_2^\dagger T_{21} d_1 \tag{1}$$

where $T_{21} = T_{12}^\dagger$. Due to inversion about the bond center and time-reversal, in the $t_{2g}$ basis we have that $T_{12} = T_{12}^\dagger = T_{12}^*$ so $T_{12}$ is real and symmetric.

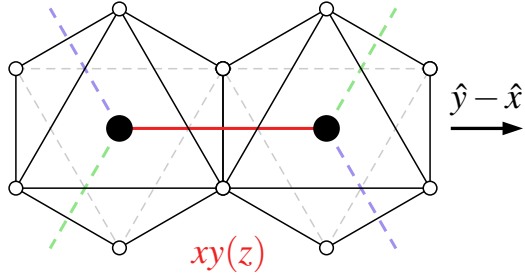

FIG. 1: The *xy(z)* bond on which we will compute the effective Hamiltonian

When only the two Ir$^{4+}$ ions and the octahedra of O$^{2-}$ ions are included the form of $T_{12}$ is constrained by symmetry. These symmetries are inversion through the bond center, as well as $C_2$ axes through the [110], [1$\bar{1}$0] and [001] axes, giving the form

$$T_{12} = \begin{pmatrix} t_1 & t_2 & 0 \\ t_2 & t_1 & 0 \\ 0 & 0 & t_3 \end{pmatrix} \tag{2}$$

with the three independent real parameters $t_1$, $t_2$ and $t_3$. The actual bond symmetry in the crystal can be lower, for example due to trigonal distortion. Within the $t_{2g}$ manifold this can introduce additional *xz* − *xy* and *yz* − *xy* hoppings such as

$$T_{12} = \begin{pmatrix} t_1 & t_2 & t_4 \\ t_2 & t_1 & t_4 \\ t_4 & t_4 & t_3 \end{pmatrix} \tag{3}$$

---
* Electronic Address: hykee@physics.utoronto.ca

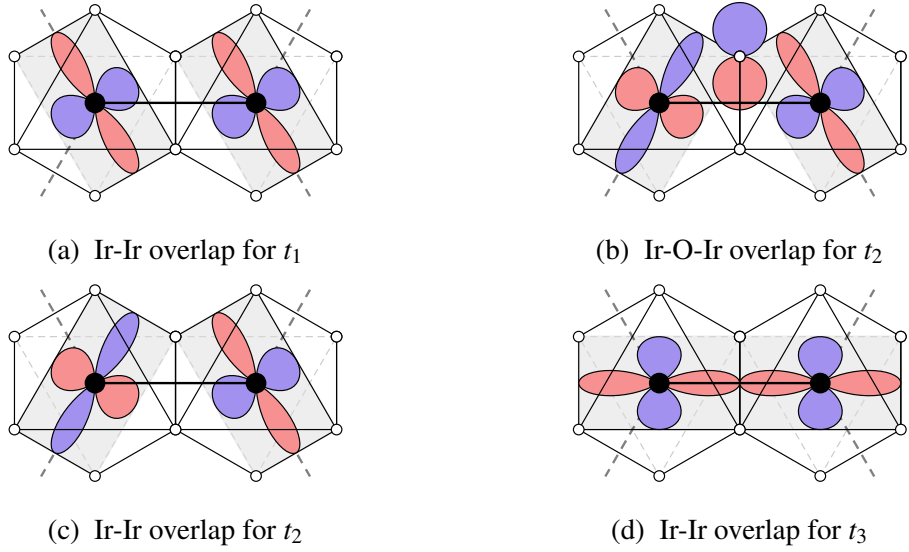

FIG. 2: Schematic visual representation of the types orbital overlap contributing to the hoppings $t_1$, $t_2$ and $t_3$ in Eq. 4a for the $xy(z)$ bond.

We will discuss two possibilities relevant for Na$_2$IrO$_3$ and Li$_2$IrO$_3$: direct overlap between the $d$ orbitals and hopping mediated through the O$^{2-}$ ions.

$$t_1 = \frac{1}{2}(t_{dd\pi} + t_{dd\delta}) \tag{4a}$$

$$t_2 = \frac{1}{2}(t_{dd\pi} - t_{dd\delta}) + \frac{t_{pd\pi}^2}{\Delta_{pd}} \tag{4b}$$

$$t_3 = \frac{1}{4}(3t_{dd\sigma} + t_{dd\delta}) \tag{4c}$$

The three parameters $t_{dd\sigma}$, $t_{dd\pi}$ and $t_{dd\delta}$ are the usual Slater-Koster parameters for direct $d$-orbital overlap, with one expecting $|t_{dd\sigma}| > |t_{dd\pi}| > |t_{dd\delta}|$. The oxygen mediated hopping is through the $t_{pd\pi}$ overlap with $\Delta_{pd}$ being the chemical potential difference between the Ir$^{4+}$ and O$^{2-}$ ions. A visual representation of these overlaps is shown in Fig. 2.

## II. STRONG COUPLING EXPANSIONS

To derive the effective Hamiltonian for the $j_{\text{eff}} = 1/2$ states we consider two forms of strong coupling expansion. The first is the conventional case, where we take $U, J_H \gg \lambda \gg t$. We consider the Kanamori Hamiltonian for the two atoms

$$H_1 + H_2 = \frac{U - 3J_H}{2}\left[(N_1 - 5)^2 + (N_2 - 5)^2\right] - 2J_H(S_1^2 + S_2^2) - \frac{J_H}{2}(L_1^2 + L_2^2) \tag{5}$$

3

where $U$ is the (screened) Coulomb interaction, $J_H$ is Hund's coupling, $N$ is the total density operator, $L$ is the total (effective) angular momentum operator and $S$ is the total spin operator.

Treating $T + T^\dagger$ as a perturbation to $H_1 + H_2$, an effective Hamiltonian within the $j_{\text{eff}} = 1/2$ subspace can be determined

$$H_{\text{eff}} \sim -\sum_{\alpha\beta}\sum_{n\neq 0}\left[\frac{\langle\alpha|T^\dagger|n\rangle\langle n|T|\beta\rangle}{E_n - E_0} + \frac{\langle\alpha|T|n\rangle\langle n|T^\dagger|\beta\rangle}{E_n - E_0}\right]|\alpha\rangle\langle\beta| \tag{6}$$

where $\alpha, \beta$ run over $j_{\text{eff}} = 1/2$ states, $\left|+\tfrac{1}{2},+\tfrac{1}{2}\right\rangle$, $\left|+\tfrac{1}{2},-\tfrac{1}{2}\right\rangle$, $\left|-\tfrac{1}{2},+\tfrac{1}{2}\right\rangle$ and $\left|-\tfrac{1}{2},-\tfrac{1}{2}\right\rangle$. The excited states of $H_1 + H_2$ are denoted by $|n\rangle$ with energies $E_n$ greater than the ground state energy $E_0$. By symmetry, the resulting spin Hamiltonian takes the form

$$H_{\text{eff}} \sim J\vec{S}_1\cdot\vec{S}_2 + K S_1^z S_2^z + \Gamma\left(S_1^x S_2^y + S_1^y S_2^x\right) + \Gamma'\left(S_1^x S_2^z + S_1^z S_2^x + S_1^y S_2^z + S_1^z S_2^y\right) \tag{7}$$

Carrying out the expansion one arrives at the following expressions for the exchange constants

$$J = \frac{4}{27}\left[\frac{6t_1(t_1 + 2t_3) - 9t_4^2}{U - 3J_H} + \frac{9t_4^2 + 2(t_1 - t_3)^2}{U - J_H} + \frac{(2t_1 + t_3)^2}{U + 2J_H}\right] \tag{8a}$$

$$\approx \frac{4}{9U}\left[(2t_1 + t_3)^2 + \frac{2J_H}{U}\left[2t_1(t_1 + 2t_3) - 3t_4^2\right]\right] + O\left(\frac{J_H^2}{U^2}\right)$$

$$K = \frac{8J_H}{9}\left[\frac{(t_1 - t_3)^2 - 3(t_2^2 - t_4^2)}{(U - 3J_H)(U - J_H)}\right] \approx \frac{8}{9U}\left(\frac{J_H}{U}\right)\left[(t_1 - t_3)^2 - 3(t_2^2 - t_4^2)\right] + O\left(\frac{J_H^2}{U^2}\right) \tag{8b}$$

$$\Gamma = \frac{8J_H}{9}\left[\frac{3t_4^2 + 2t_2(t_1 - t_3)}{(U - 3J_H)(U - J_H)}\right] \approx \frac{8}{9U}\left(\frac{J_H}{U}\right)\left[3t_4^2 + 2t_2(t_1 - t_3)\right] + O\left(\frac{J_H^2}{U^2}\right) \tag{8c}$$

$$\Gamma' = -\frac{8J_H}{9}\left[\frac{t_4(t_1 - t_3 - 3t_2)}{(U - 3J_H)(U - J_H)}\right] \approx -\frac{8}{9U}\left(\frac{J_H}{U}\right)[t_4(t_1 - t_3 - 3t_2)] + O\left(\frac{J_H^2}{U^2}\right) \tag{8d}$$

The full nearest neighbour spin model can be obtained using lattice symmetries to generate the remaining bonds. The expressions given in Eqs. 3-5 of the main text can be obtained from Eq. 8 by setting $t_4 = 0$.

A similar calculation can be carried out taking the limit $U, \lambda \gg J_H \gg t$. Here the atomic Hamiltonian includes only the Coulomb interaction $U$ and spin-orbit coupling $\lambda$. The contributions proportional to $J_H$ can then be included in the eigenstates and energies using (degenerate)

4

perturbation theory, and $H_{\text{eff}}$ evaluated as above. This gives expressions

$$J = \frac{4}{27}\left[\frac{(2t_1+t_3)^2(4J_H+3U)}{U^2} - \frac{8J_H\left(9t_4^2+2(t_1-t_3)^2\right)}{(2U+3\lambda)^2}\right] \tag{9a}$$

$$K = \frac{32J_H}{9}\left[\frac{(t_1-t_3)^2-3(t_2^2-t_4^2)}{(2U+3\lambda)^2}\right] \tag{9b}$$

$$\Gamma = \frac{32J_H}{9}\left[\frac{3t_4^2+2t_2(t_1-t_3)}{(2U+3\lambda)^2}\right] \tag{9c}$$

$$\Gamma' = -\frac{32J_H}{9}\left[\frac{t_4(t_1-t_3-3t_2)}{(2U+3\lambda)^2}\right] \tag{9d}$$

up to corrections of order $O(J_H^2/\lambda^2)$ and $O(J_H^2/U^2)$. Note that the dependence of these exchanges on the hoppings is nearly identical in both perturbation theories.

## III. DERIVATION OF Γ TERM

To unambiguously show the existence of the Γ term in the strong-coupling Hamiltonian, we derive the expression for Γ explicitly. We will work in $U, J_H \gg \lambda \gg t$ limit to connect with the equations shown in the main text. The coefficient Γ can then be determined from the matrix element

$$\Gamma = 2i\left\langle -\tfrac{1}{2}, -\tfrac{1}{2}\left|H_{\text{eff}}\right|+\tfrac{1}{2}+\tfrac{1}{2}\right\rangle \tag{10}$$

where $H_{\text{eff}}$ is the two-site effective Hamiltonian for a $xy(z)$ bond. To simplify notation we use the particle-hole mapping $d^5 \to d^1$, mapping the Hamiltonian of Eq. 5 to

$$H_0 = H_1 + H_2 = \frac{U-3J_H}{2}\left[(N_1-1)^2+(N_2-1)^2\right] - 2J_H\left(S_1^2+S_2^2\right) - \frac{J_H}{2}\left(L_1^2+L_2^2\right) \tag{11}$$

Since particle-hole conjugation effectively maps $\left|+\tfrac{1}{2}\right\rangle \leftrightarrow \left|-\tfrac{1}{2}\right\rangle$ we see that we are interested in $\left\langle +\tfrac{1}{2},+\tfrac{1}{2}\left|H_{\text{eff}}\right|-\tfrac{1}{2}-\tfrac{1}{2}\right\rangle = \left\langle -\tfrac{1}{2},-\tfrac{1}{2}\left|H_{\text{eff}}\right|+\tfrac{1}{2}+\tfrac{1}{2}\right\rangle^*$ in this new basis. The eigenstates of $H_1$ or $H_2$ are eigenstates of $N, L^2, L_z, S^2$ and $S_z$ at each site, which we will denote as sets using terms symbols $^{2S+1}L$. Each individual eigenstates will be denoted as $\left|^{2S+1}L, M_L, M_S\right\rangle_a$ with the total number $N$ understood from context and $a = 1, 2$ denoting the site index. The ground state is an $N = 1$ six-fold degenerate $^2P$ state with energy $-5J_H/2$. Since the perturbation $T$ moves an electron from site 2 to site 1 we need then only consider $N_1 = 2$ and $N_2 = 0$ states. While the $N = 0$ state is a trivial $^1S$ state, anti-symmetric $N = 2$ states can be formed by $^1S, ^3P$ and $^1D$ terms giving three relevant

5

excitation energies

$$E_2(^1S) + E_0(^1S) - 2E_1(^2P) = U + 2J_H \tag{12a}$$

$$E_2(^3P) + E_0(^1S) - 2E_1(^2P) = U - 3J_H \tag{12b}$$

$$E_2(^1D) + E_0(^1S) - 2E_1(^2P) = U - J_H \tag{12c}$$

where we have denoted the energy of the $N$-electron term $^{2S+1}L$ as $E_N(^{2S+1}L)$. The most straightforward way to compute $\Gamma$ is to first decompose $T\left|+\tfrac{1}{2}\right\rangle_1 \left|+\tfrac{1}{2}\right\rangle_2$ into eigenstates of $H_1 + H_2$. Break $T$ into two parts $T = T_1 + T_2$ with

$$T_1 = \sum_\sigma \left[ t_1 \left( d^\dagger_{1,xz,\sigma} d_{2,xz,\sigma} + d^\dagger_{1,yz,\sigma} d_{2,yz,\sigma} \right) + t_3 \sum_\sigma d^\dagger_{1,xy,\sigma} d_{2,xy,\sigma} \right] \tag{13a}$$

$$T_2 = t_2 \sum_\sigma \left( d^\dagger_{1,xz,\sigma} d_{2,yz,\sigma} + d^\dagger_{1,yz,\sigma} d_{2,xz,\sigma} \right) \tag{13b}$$

Recall the definition of the $j_{\text{eff}} = 1/2$ states as

$$\left|+\tfrac{1}{2}\right\rangle_a = \sqrt{\tfrac{1}{3}} \left( d^\dagger_{a,yz,\downarrow} + i d^\dagger_{a,xz,\downarrow} + d^\dagger_{a,xy,\uparrow} \right) |0\rangle_a = \left[ \sqrt{\tfrac{2}{3}} d^\dagger_{a,+,\downarrow} - i \sqrt{\tfrac{1}{3}} d^\dagger_{a,0,\uparrow} \right] |0\rangle_a \tag{14a}$$

$$\left|-\tfrac{1}{2}\right\rangle_a = \sqrt{\tfrac{1}{3}} \left( d^\dagger_{a,yz,\uparrow} - i d^\dagger_{a,xz,\uparrow} - d^\dagger_{a,xy,\downarrow} \right) |0\rangle_a = \left[ \sqrt{\tfrac{2}{3}} d^\dagger_{a,-,\uparrow} + i \sqrt{\tfrac{1}{3}} d^\dagger_{a,0,\downarrow} \right] |0\rangle_a \tag{14b}$$

where $a = 1, 2$ and we have defined the $l_{\text{eff}} = 1$ operators

$$d^\dagger_{a,+,\sigma} = \sqrt{\tfrac{1}{2}} \left( d^\dagger_{a,yz,\sigma} + i d^\dagger_{a,xz,\sigma} \right) \tag{15a}$$

$$d^\dagger_{a,0,\sigma} = i d^\dagger_{a,xy,\sigma} \tag{15b}$$

$$d^\dagger_{a,-,\sigma} = \sqrt{\tfrac{1}{2}} \left( d^\dagger_{a,yz,\sigma} - i d^\dagger_{a,xz,\sigma} \right) \tag{15c}$$

In this form we can now act each part of $T$, decomposing into the excited states as we go

$$\begin{aligned}
T_1 \left|+\tfrac{1}{2}\right\rangle_1 \left|+\tfrac{1}{2}\right\rangle_2 &= \left(\tfrac{t_1 - t_3}{3}\right) \left( d^\dagger_{1,yz,\downarrow} + i d^\dagger_{1,xz,\downarrow} \right) d^\dagger_{1,xy,\uparrow} |0\rangle_1 |0\rangle_2 \\
&= -i \left(\tfrac{t_1 - t_3}{3}\right) \left[ |+\downarrow, 0\uparrow\rangle_1 - |0\uparrow, +\downarrow\rangle_1 \right] |0\rangle_2 \\
&= -i \left(\tfrac{t_1 - t_3}{3}\right) \left[ |^3P, +1, 0\rangle_1 - |^1D, +1, 0\rangle_1 \right] |0\rangle_2
\end{aligned} \tag{16a}$$

$$\begin{aligned}
T_2 \left|+\tfrac{1}{2}\right\rangle_1 \left|+\tfrac{1}{2}\right\rangle_2 &= \tfrac{t_2}{3} \left[ 2 d^\dagger_{1,xz,\downarrow} d^\dagger_{2,yz,\downarrow} + i \left( d^\dagger_{1,yz,\downarrow} - i d^\dagger_{1,xz,\downarrow} \right) d^\dagger_{1,xy,\uparrow} \right] |0\rangle_1 |0\rangle_2 \\
&= \tfrac{t_2}{3} \left[ -2i \left( |+\downarrow, -\downarrow\rangle - |-\downarrow, +\downarrow\rangle \right) + \left( |-\downarrow, 0\uparrow\rangle - |0, \uparrow, -\downarrow\rangle \right) \right] |0\rangle_2 \\
&= \tfrac{t_2}{3} \left[ -2\sqrt{2} i \left|^3P, 0, -1\right\rangle - \left|^3P, -1, 0\right\rangle - \left|^1D, -1, 0\right\rangle \right] |0\rangle_2
\end{aligned} \tag{16b}$$

6

To compute the matrix element we also need $T\left|-\frac{1}{2}\right\rangle_1 \left|-\frac{1}{2}\right\rangle_2$ related to the states shown in Eq. 16 by the action of time-reversal

$$T_1 \left|-\tfrac{1}{2}\right\rangle_1 \left|-\tfrac{1}{2}\right\rangle_2 = -i\left(\frac{t_1 - t_3}{3}\right)\left[|^3P, -1, 0\rangle_1 - |^1D, -1, 0\rangle_1\right]|0\rangle_2 \tag{17a}$$

$$T_2 \left|-\tfrac{1}{2}\right\rangle_1 \left|-\tfrac{1}{2}\right\rangle_2 = \frac{t_2}{3}\left[-2\sqrt{2}i\,|^3P, 0, +1\rangle + |^3P, +1, 0\rangle + |^1D, +1, 0\rangle\right]|0\rangle_2 \tag{17b}$$

where we have used that the $\left|^{2S+1}L, M_L, M_S\right\rangle$ states pick up an additional sign $(-1)^{L+S+M_L+M_S}$ under time-reversal. In this form we can simply read off the matrix elements

$$\begin{aligned}
\left\langle -\tfrac{1}{2}, -\tfrac{1}{2} \,|H_{\text{eff}}|\, +\tfrac{1}{2} + \tfrac{1}{2}\right\rangle &= 4i\left(\frac{t_2(t_1 - t_3)}{9}\right)\left[\frac{1}{U - 3J_H} - \frac{1}{U - J_H}\right] \\
&= \frac{8iJ_H}{9}\left[\frac{t_2(t_1 - t_3)}{(U - 3J_H)(U - J_H)}\right]
\end{aligned} \tag{18}$$

Thus

$$\Gamma = \frac{16J_H}{9}\left[\frac{t_2(t_1 - t_3)}{(U - 3J_H)(U - J_H)}\right] \tag{19}$$

consistent with the expressions in the Eq. 5 of the main text and with Eq. 8c when $t_4 = 0$.



\end{document}